\documentclass[aps,twocolumn,amsfonts,amssymb,amsmath,pra,showpacs,superscriptaddress]{revtex4}

\usepackage{graphicx}
\usepackage{epsfig}
\usepackage{amsmath}
\usepackage{amssymb}
\usepackage{bm}

\newcommand{\comment}[1]{}
\newcommand{\e}{\mbox{e}}
\newcommand{\I}{\mbox{i}}

\begin{document}

\title{Conditional preparation of arbitrary atomic Dicke states}

\author{Karel Lemr}
\affiliation{Joint Laboratory of Optics of Palack\'y University and Institute of Physics of Academy of Sciences of the Czech Republic, 17. listopadu 50A, 77200 Olomouc, Czech Republic}\
\author{Jarom\'{\i}r Fiur\'a\v{s}ek}
\affiliation{Department of Optics, Palack\'y University, 17. listopadu 50, 77200 Olomouc, Czech Republic}

\begin{abstract}
We propose an experimentally accessible procedure for conditional preparation of highly non-classical states of collective spin of an atomic ensemble. 
The quantum state engineering is based on a combination of QND interaction between atoms and light previously prepared in a non-Gaussian state using photon subtraction from squeezed vacuum beam, homodyne detection on the output light beam, and a coherent displacement of atomic state. The procedure is capable of non-deterministic preparation of a wide class of superpositions of atomic Dicke states. We present several techniques to optimize the performance of the protocol and maximize the trade-off between fidelity of prepared state and success probability of the scheme.

\end{abstract}


\maketitle

\section{Introduction}

Laws of quantum physics enable to process and transmit information in ways that would be impossible or very difficult to achieve classically. Prime example is the unconditionally secure quantum key distribution that is already approaching the stage of commercial applications \cite{Scarani08}. Most of the developed quantum communication protocols employ light as a carrier and processing medium. Quantum information needs however not only to be transmitted and processed, but it requires also storage facilities. Storage of quantum information can not be accomplished using classical means, but it requires a special - quantum - memory. Quantum memory
is indispensable for construction of quantum repeaters \cite{Briegel98,Duan01} that combine it with entanglement distillation and swapping to efficiently distribute entanglement over lossy and noisy quantum channels. A very promising medium for quantum memory are ensembles of atoms  trapped in the electromagnetic field \cite{Matsukevich04,Chaneliere05,Choi08,Yuan08,Smith08,Appel08} or held in a glass cell 
\cite{Julsgaard01,Schori02,Julsgaard04,Cviklinski08}.

Ultimately, quantum memory should allow to store a quantum state of one or several modes of the electromagnetic field, and not only states of single photons. Particularly promising for this purpose appears to be the off-resonant quantum non-demolition (QND) coupling between a collective (pseudo)-spin of the atomic ensemble and the polarization of light beam \cite{Kuzmich98,Kuzmich00,Duan00,Hammerer08}. This setting involves an auxiliary coherent laser beam that mediates the coupling between an atomic ensemble and a light mode whose quantum state should be stored  to the atoms. Using large atomic ensembles and strong auxiliary laser beams leads to a collective enhancement of the atoms-light coupling. Consequently, sufficiently strong interaction can be achieved even without a cavity just by a single passage of light through the atomic sample \cite{Julsgaard01,Julsgaard04,Hammerer08,Dantan05b}, which greatly simplifies the experiment. 
The QND coupling has been explored in several landmark experiments to entangle states of two atomic ensembles \cite{Julsgaard01}, to store quantum state of a light beam onto the collective spin of  atoms \cite{Julsgaard04}, and to teleport quantum state of light  onto atoms \cite{Sherson06}. Besides quantum information processing applications, the QND coupling can be also used to generate spin squeezed state of the atomic ensemble that can enhance precision of atomic clocks \cite{Smith08,Appel08}.

Collective spin degree of freedom  of atomic ensemble can be described by collective atomic spin operators $\hat{J}_x$, $\hat{J}_y$ and $\hat{J}_z$. By making the expectation value of one of these operators sufficiently large, the other two manifest similar algebraic properties as quadrature operators of light.
The QND coupling between atoms and light can then  be  described by linear input-output transformations of the effective quadrature operators of atoms and light \cite{Duan00,Kuzmich00,Hammerer08}. This greatly simplifies  theoretical analysis of atomic memory operation \cite{Fiurasek03,Molmer04,Hammerer05,Muschik06,Fiurasek06,Sherson06b} but also somewhat restricts possible manipulations with memory. 
Experimentally easily accessible coherent and squeezed light beams are described by  Gaussian Wigner functions
and all such states are referred to as Gaussian states.  Using Gaussian light states and QND coupling, possibly combined with homodyne detection on output light and feedback, we can implement only Gaussian operations on the atomic memory and prepare only Gaussian states in the memory starting from initial Gaussian state. 

Certain applications, notably continuous-variable entanglement distillation, however require non-Gaussian operations \cite{Eisert02,Giedke02,Fiurasek02,Browne03,Eisert04,Hage08,Dong08}.
It is therefore highly desirable to investigate schemes for implementation of non-Gaussian operations and filters on the quantum-memory state and devise procedures for preparation of arbitrary highly nonclassical states of atomic quantum memory. 
Previously, scheme for generation of a superposition of two coherent spin states of an atomic ensemble has been proposed \cite{Massar03} and a protocol for probabilistic noise-free upload of single-photon and Schr\"{o}dinger cat-like states into atomic memory has been suggested \cite{Filip08}.

In this paper, we propose a scheme for conditional preparation of arbitrary coherent superpositions of atomic Dicke states $|n\rangle$,
	\begin{equation}
	\label{eq:objectif}
	|\psi_{\mbox{\footnotesize{target}}}\rangle = \sum^{N}_{n=0} c_n |n\rangle.
	\end{equation}
Let $N_A$ denotes the number of atoms in the ensemble. The quantum memory typically exploits coherence between two atomic Zeeman or hyperfine levels $|\uparrow\rangle$ and $|\downarrow\rangle$. The Dicke state
 $|n\rangle$ is then defined as a fully symmetric state of all $N_A$ atoms with $n$ atoms in state $|\uparrow\rangle$ and $N_A-n$ atoms in state 
$|\downarrow\rangle$. In the limit of large $N_A$, the states $|n\rangle$ become formally equivalent to the $n$-photon Fock states of light. 

The general idea behind our protocol is to manipulate the atomic state by QND interaction with light that has been previously prepared in a specific highly non-classical quantum state. This technique allows us to employ QND interaction to implement operations on atomic ensemble that are not easy to perform directly on atoms, but are more feasible on light.
In particular, we show that by using light beam prepared in photon-subtracted squeezed vacuum state  \cite{Ourjoumtsev06,Nielsen06,Wakui06} we could implement an operation on the atomic memory that is similar to single-photon subtraction/addition. 

This elementary non-Gaussian operation can be combined with coherent displacements of atomic state and repeated several times to  conditionally generate a wide class of superpositions (\ref{eq:objectif}). The resulting scheme is analogous to protocols for generation of arbitrary superpositions of Fock states of traveling light beams by repeated photon addition or subtraction \cite{Dakna99,Fiurasek05}, but it exhibits important differences due to the QND coupling and the need for conditioning on the outcomes of homodyne detection on the output light beam. We analyze in detail influence of various relevant experimental parameters on the performance of the protocol and show how it can be optimized in order to maximize the fidelity of generated state for a given probability of success.

 	\begin{figure}[!t!]
	\begin{center}
	\scalebox{1.25}{\includegraphics{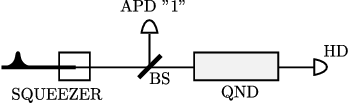}}
	\caption{Atoms-light interaction setup. Light is prepared in squeezed vacuum state in the squeezer and then a tiny portion of the light beam is reflected from an unbalanced  beam splitter BS and impinges on single-photon detector APD. Click of APD heralds subtraction of single-photon from the squeezed beam.  Light in such state accompanied by orthogonally polarized strong coherent beam interacts with atomic ensemble. Afterwords, measurement of $p_L$ quadrature is performed on output light beam using homodyne detector (HD).}
	\label{fig:qnd}
	\end{center}
	\end{figure}
	
The rest of the paper is organized as follows. In Section~II we will review the basics of QND coupling between light and atoms and derive a general formula for the operation performed on atoms.  In Sec.~III we explain how the elementary non-Gaussian operation can 
be combined with magnetic-field induced atomic displacement operator in order to generate various superpositions of states $|0\rangle$ and $|1\rangle$. In Section~IV, the procedure is generalized to multiple repeated application of non-Gaussian operation and coherent displacement, which allows to generate wide class of superpositions (\ref{eq:objectif}). Results of numerical simulations and strategies for optimization of the protocol are presented 
in Sec.~V. In Section~VI we show that by controlling which atomic quadrature couples to light we can in principle generate arbitrary complex superpositions (\ref{eq:objectif}). In Sec.~VII we discuss an alternative procedure for direct single-step preparation of atomic state by using a specifically prepared highly non-classical state of light, which is then imprinted onto atoms. Finally, Section~VIII contains a brief summary and conclusions.

\section{Atoms-light interaction}

Our scheme for preparation of highly non-classical states of the atomic ensemble 
is based on the off-resonant quantum non-demolition interaction between light beam and collective atomic spin. 
The light beam propagating along $z$ axis consists of a strong coherent vertically polarized mode and a horizontally polarized mode prepared in a pure quantum state $|\phi_L\rangle$. The atoms are prepared by optical pumping in a coherent spin state with all spins pointing along the $x$ axis such that the $x$-component of the collective atomic spin $\bm{J}$ attains a macroscopic value, $\langle \hat{J}_x\rangle = F N_A$, where $N_A$ is the total number of atoms in the cloud. Under this condition we can replace operator $\hat{J}_x$ with its mean value in the commutation relations $[\hat{J}_y,\hat{J}_z]=i\hat{J}_x$ and define the effective atomic quadratures \cite{Hammerer08}
	\begin{equation}
	\label{eq:atomic_quadratures}
	\hat{x}_{A} = \frac{\hat{J}_y}{\sqrt{2\langle\hat{J}_x\rangle}}, \quad \hat{p}_{A} = \frac{\hat{J}_z}{\sqrt{2\langle\hat{J}_x\rangle}},
	\end{equation}
satisfying canonical commutation relations
	\begin{equation}
	\label{eq:commutation_relations}
	[\hat{x}_{A}, \hat{p}_{A}] = \frac{i}{2}.
	\end{equation}
Similarly, the quadratures of the horizontally polarized light mode are defined such that   $[\hat{x}_L,\hat{p}_L]=i/2$ holds. 
With this normalization the wave-function of vacuum reads $\phi_{\mathrm{vac}}(x)=(2/\pi)^{1/4}\exp(-x^2)$, which facilitates further calculations.
Using quadrature operators  one can write down the effective 
QND interaction Hamiltonian between light and atoms in the form of \cite{Kuzmich98,Kuzmich00,Hammerer08}
	\begin{equation}
	\label{eq:qnd_hamilton}
	\hat{H}_{\mathrm{QND}} = \hbar \bar{\kappa}  \hat{x}_L \hat{x}_A,
	\end{equation}
where $\bar{\kappa}$ is the interaction constant. 

Suppose that before the interaction atoms and light are in the initial pure states $|\phi_A\rangle$ and $|\phi_L\rangle$, respectively. Light beam passes through the atoms and is subsequently subjected to homodyne detection of the $\hat{p}_L$ quadrature. The operation performed on the atomic state corresponding to a particular measurement outcome $p_L$  can be expressed as
	\begin{equation}
	\label{eq:qnd_procedure_general}
	\hat{\Theta} = \langle p_L| \exp{\left (- \I \hat{H}_{\mathrm{QND}} t / \hbar \right )} |\phi_L\rangle,
	\end{equation}
where $|p_L\rangle$ is the eigenstate of $\hat{p}_L$ with eigenvalue $p_L$.
	Working in the $x$-representation and using the fact that $\langle p_L|x_L\rangle=\frac{1}{\sqrt{\pi}}e^{-i2x_L p_L}$ we obtain
	\begin{equation}
	\label{eq:qnd_procedure_PHI}
	\hat{\Theta} = \frac{1}{\sqrt{\pi}} \int_{-\infty}^{\infty} \e^{-2i x_L p_L} \e^{-2 i \kappa \hat{x}_A x_L} \phi_L(x_L) \mathrm{d}x_L,	
	\end{equation}
where $\kappa =\bar{\kappa}t /2$. This formula can be written in a concise form
\begin{equation}
\label{eq:theta_fourier}
\hat{\Theta} = \Phi(\kappa \hat{x}_A + p_L),
\end{equation}
where $\Phi(p_L)$ denotes the Fourier transform of the wave function $\phi_L(x_L)=\langle x_L|\phi_L\rangle$.

Formula (\ref{eq:theta_fourier}) reveals that in order to apply a non-Gaussian operation on the atomic state we
need non-Gaussian state of light beam $|\phi_L\rangle$. Recently, highly non-Gaussian states of traveling light beam exhibiting negative Wigner function have been generated by photon subtraction from squeezed vacuum \cite{Ourjoumtsev06,Nielsen06,Wakui06}.   As schematically shown in Fig.~\ref{fig:qnd}, squeezed light beam is generated in an optical parametric amplifier and then impinges on a highly unbalanced  beam splitter BS with nearly unit transmissivity. 
Click of the single-photon detector APD indicates with high probability  a subtraction of single photon because probability that two or more photons are reflected is negligible. 
The resulting conditional transformation can be described by the action of annihilation operator 
on the initial squeezed state,
	\begin{equation}
	\label{eq:qnd_a_action}
	\hat{S}_L(r) |0_L\rangle \longrightarrow \hat{a}_L \hat{S}_L(r) |0_L\rangle,
	\end{equation}
where $\hat{S}_L(r) = \exp \left [\frac{r}{2}(\hat{a}_L^{\dag 2}-\hat{a}_L^2 )\right ]$ denotes the squeezing operator and $|0_L\rangle$ is the vacuum state.
The success probability of the whole procedure depends on the efficiency of the detector and also on the reflectivity of the beam splitter. One may increase the success probability of the operation by using more reflective beam splitter. This will however result in higher probability of reflection of two photons and therefore in lower fidelity of the prepared state. In the limit of very low reflectance of BS, the photon-subtracted squeezed state is formally equivalent to squeezed single-photon Fock state and its normalized wave-function reads 
\begin{equation}
\phi_L(x_L)=\left(\frac{2}{\pi}\right)^{1/4} e^{3r/2}\, 2x_L\exp\left(- e^{2r} x_L^2\right).
\end{equation}
Performing Fourier transformation we obtain the resulting operator
	\begin{equation}
	\label{eq:qnd_operator}
	\hat{\Theta}_S(p_L) = \mathcal{N}(\hat{x}_A+p_L/\kappa) \exp{\left [- \epsilon(\hat{x}_A+p_L/\kappa)^2 \right ]},
	\end{equation}
where $\epsilon = \kappa^2\e^{-2r}$ and $\mathcal{N}$ stands for normalization constant,
	\begin{equation}
	\label{eq:qnd_N}
	\mathcal{N} = 2\kappa\left(\frac{2}{\pi}\right)^{1/4} e^{-3r/2}.
	\end{equation}
The operator $\hat{\Theta}_S$ is the key essence of our quantum state engineering scheme. 
One can easily verify that action of $\hat{\Theta}_S(0)$ on vacuum atomic state $|0_A\rangle$ leads to the final atomic state whose wave function reads
	\begin{equation}
	\label{eq:1_wave}
	\langle x_A|\hat{\Theta}_S(0) |0_A\rangle =\left(\frac{2}{\pi}\right)^{1/4} \mathcal{N}  x_A \exp{\left [- (\epsilon+1) x_A^2 \right ]}.
	\end{equation}
We thus obtain a highly nonclassical  squeezed  state $|1\rangle$ in the atomic memory \cite{Filip08}.
We can see that the $\hat{\Theta}_S(0)$ operator acts simultaneously as a combination of creation and annihilation operator and a squeezing operator. As shown in next section, this property  can be exploited to generate (squeezed) superpositions of higher Dicke states.

\section{Combination with displacement operator}

Displacement operator $\hat{D}(\alpha)=\exp(\alpha\hat{a}^\dagger-\alpha^\ast\hat{a})$ can be easily implemented on atomic ensemble by application of magnetic field resulting in a tiny rotation of the collective atomic spin. We propose to combine displacement operator and the $\hat{\Theta}_S$ operator to generate various superpositions of states $|0\rangle$ and $|1\rangle$. In this and the following section we will assume conditioning on the measurement outcome $p_L=0$ corresponding to operation $\hat{\Theta}_S(0)$. 
In practice, we have to use a finite acceptance window for the measurement outcomes 
in order to achieve finite success probability. The effect of the width of the acceptance window on fidelity of generated state will be discussed in Sec. V.  

Consider a sequence of coherent displacement followed by non-Gaussian operation $\hat{\Theta}_S(0)$ and another displacement. The entire operation on the atomic ensemble takes the form
	\begin{equation}
	\label{eq:1_sequence}
	\hat{D}(\alpha_2) \hat{\Theta}_S(0) \hat{D}(\alpha_1).
	\end{equation}
The displacement operator $\hat{D}(\alpha_2)$ can be propagated to the right using 
$\hat{D}(\alpha_j) \hat{x}=(\hat{x}-a_j)\hat{D}(\alpha_j) $, where $\alpha_j=a_j+ib_j$ and $a_j$, $b_j$ are real. We obtain
\begin{widetext}
	\begin{equation}
	\label{eq:1_result_operator}
	\hat{D}(\alpha_2) \hat{\Theta}_S(0) \hat{D}(\alpha_1) 
	=e^{i\Im(\alpha_2\alpha_1^\ast)} (\hat{x}_A - a_2) 
	\exp \left [ -\epsilon (\hat{x}_A-a_2)^2 \right ] \hat{D}(\alpha_1 + \alpha_2).
	\end{equation}
Assuming the atoms are initially in the effective vacuum state $|0\rangle$ the transformation (\ref{eq:1_sequence}) prepares the atoms in a state
	\begin{equation}
	\label{eq:1_result_wave_n1}
	\psi_\mathrm{A,out}(x_A) \propto (x_A - a_2) \exp \left [ -(\epsilon + 1) x_A^2 + 2\left (a_1+(\epsilon+1) a_2\right ) x_A + 2i x(b_1+b_2) \right ],
	\end{equation}
\end{widetext}	
where an unimportant constant has been neglected.
Wave function (\ref{eq:1_result_wave_n1}) represents a squeezed superposition of vacuum and $|1\rangle$ state provided that its Gaussian part is centered on origin. This can  be accomplished simply by putting
	\begin{eqnarray}
	\label{eq:1_conditions}
	&b_1  =  b_2 = 0, & \nonumber \\
	&a_1  =  - (\epsilon + 1) a_2.&
	\end{eqnarray}
By changing the free parameter $a_2$, one can easily modify relative amplitudes of states $|0\rangle$ and $|1\rangle$ in the superposition (\ref{eq:1_result_wave_n1}) as long as they remain real, i.e. without relative phase shift. We can thus prepare in the atomic memory arbitrary squeezed state $S(s)(c_0|0\rangle+c_1|1\rangle)$, where $c_0,c_1$ are real and $s=\frac{1}{2}\log(\epsilon+1)$. In current experiments, $\kappa \leq 1$ holds. If the light beam is squeezed in the amplitude quadrature, $r>0$, then $\epsilon \ll 1$ and the resulting squeezing of the state in atomic 
memory is small and could be neglected. However, if the light beam is squeezed in phase quadrature, $r<0$, then  the resulting atomic squeezing may become significant. We will see in Sec. V that choosing $r<0$ may be advantageous since it can provide better trade-off between success probability of the protocol and fidelity of the generated state.

	\begin{figure}[!b!]
	\begin{center}
	\includegraphics[width=\linewidth]{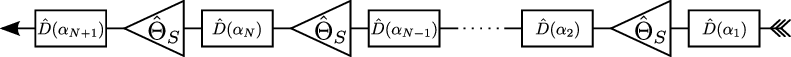}
	\caption{Schematized multiple interaction scheme for generation of superpositions of first $N+1$ Dicke states. The procedure consists of sequence of $N$ non-Gaussian operations combined  with $N+1$ coherent displacements of the atomic state, as described by Eq. (\ref{eq:multi_sequence}).}
	\label{fig:qnd_repeat}
	\end{center}
	\end{figure}

The fact that the target state $c_0|0\rangle+c_1|1\rangle$
is squeezed is not necessarily an obstacle and, in fact, it can be desirable in certain cases. For instance, the squeezed  state $|1\rangle$ very well approximates  
superposition of two coherent states $|\alpha\rangle-|-\alpha\rangle$, which is often referred to as Schr\"{o}dinger cat-like state and is a valuable resource for quantum information processing. The squeezing can be removed by the anti-squeezing operation $S(-s)$ performed on the atomic state. This can be accomplished by means of an auxiliary light beam that interacts with the atomic state several times either sequentially \cite{Fiurasek03} or simultaneously \cite{Fiurasek06}. If $\epsilon<1$ then one could also pre-squeeze the atomic ensemble before state preparation to compensate for the effects of squeezing. Initializing the atomic memory in squeezed state $\psi_{A,\mathrm{in}}(x_A)\propto\exp[-(1-\epsilon)x_A^2]$ would yield the desired superposition of Dicke states without any parasitic squeezing. 
Note that in order to have the Gaussian part of the wave-function centered on vacuum, we should now set $a_1=-a_2/(1-\epsilon)$.
The initial squeezing of the state of atomic ensemble can be accomplished by  QND measurement of the $p_A$ quadrature followed by coherent displacement of the atomic state proportional to measurement outcome \cite{Julsgaard01,Hammerer08}.

\section{Multiple interaction scheme}

By repetition of the basic sequence (\ref{eq:1_sequence}) one is able to prepare superpositions of higher Dicke states. Such generalized operation acting on initial atomic vacuum state reads,
	\begin{equation}
	\label{eq:multi_sequence}
	\hat{D}(\alpha_{N+1}) \hat{\Theta}_S(0) \hat{D}(\alpha_{N}) \hat{\Theta}_S(0) 
	\ldots \hat{D}(\alpha_2) \hat{\Theta}_S(0) \hat{D}(\alpha_1)|0\rangle
	\end{equation}
and is schematized in Fig.~\ref{fig:qnd_repeat}. Similar calculation as that leading to Eq. (\ref{eq:1_wave}) gives us the resulting atomic wave function. Defining
	\begin{equation}
	\label{eq:multi_displacements}
	\tilde{\alpha}_k = \sum_{j=k}^{N+1} \alpha_j
	\end{equation}
and decomposing these cumulative displacements into real and imaginary parts, $\tilde{\alpha}_k=\tilde{a}_k+i\tilde{b}_k$, the wave function  reads
	\begin{eqnarray}
	\label{eq:multi_wave}
	\psi_A(x_A) &\propto& \prod_{k=2}^{N+1} (x_A - \tilde{a}_k) 
	\exp\left (-\epsilon N x_A^2  +	2 \epsilon \sum_{k=2}^{N+1} \tilde{a}_k x_A \right)
	\nonumber \\
	& &\times \exp\left[- \epsilon \sum_{k=2}^{N+1} \tilde{a}_k^2 + 2 i  \tilde{b}_1 x_A -
	(x_A -  \tilde{a}_1)^2 \right ]. \nonumber \\
	\end{eqnarray}
Inspired by Eq. (\ref{eq:1_conditions}), one can always set
	\begin{eqnarray}
	\label{eq:multi_conditions}
	 b_j & = & 0, \quad \forall j, \nonumber \\
	\tilde{a}_1 + \epsilon \sum_{k=2}^{N+1} \tilde{a}_k & = & 0.
	\end{eqnarray}
The Gaussian part of the wave function (\ref{eq:multi_wave}) then becomes centered on origin,
	\begin{equation}
	\label{eq:multi_wavefce_centered}
	\psi_A(x_A) \propto \prod_{k = 2}^{N+1} \left (x_A - \tilde{a}_{k} \right ) \exp{\left[-\left(\epsilon N+1\right) x_A^2\right]}.
	\end{equation}
Wave function of any Dicke state $|n\rangle$ is composed of product of a Hermite polynomial of $n$th order $H_n(\sqrt{2}x)$ and a Gaussian. Therefore the wave function of finite superposition of Dicke states (\ref{eq:objectif}) can be expressed as a product of a Gaussian and a polynomial with degree equal to the highest Dicke state in the target superposition. 
The polynomial part of the wave function $\psi_{\mathrm{target}}$ can be decomposed into 
a factorized form,
	\begin{equation}
	\label{eq:multi_fact_form}
	\psi_{\mathrm{target}}(x) \propto \mathrm{e}^{-x^2}\sum^{N}_{n=0} c_n \frac{H_n(\sqrt{2}x)}{\sqrt{2^n n!}} \propto  \mathrm{e}^{-x^2} \prod_{j=1}^{N} (x - R_j).
	\end{equation}
By comparison with the wave function that results from multiple interaction scheme (\ref{eq:multi_wavefce_centered}), we see that we are able to engineer a squeezed version of the desired state (\ref{eq:multi_fact_form}) just by setting correctly displacement parameters $\alpha_j$. We need to find roots $R_j$ of the polynomial $\sum_{n=0}^N c_n H_n(\sqrt{2}x)/\sqrt{2^n n!}$ and set $b_j=0$ and 
	\begin{eqnarray}
	\label{eq:multi_roots}
	  a_{N+1} & = & \frac{R_{N}}{\sqrt{N \epsilon +1}}, \nonumber \\
	 a_{k} & = &\frac{ R_{k-1} - R_{k}}{\sqrt{N\epsilon+1}}, \quad k = 2,\ldots, N, \nonumber \\
	a_{1} & = & \frac{1}{\sqrt{N\epsilon+1}}\left(- R_{1} - \epsilon \sum_{k=1}^{N} R_k\right).
	\end{eqnarray}
Note that this formula must yield real $a_j$. Our protocol is therefore capable of preparation of any superposition of Dicke states as long as all roots $R_j$ remain real. In particular, we can prepare arbitrary (squeezed) Dicke state $|n\rangle$, because all roots of Hermite polynomial $H_n(x)$ are real.

	\begin{figure}[!t!]
	\begin{center}
	\scalebox{1.0}{\includegraphics{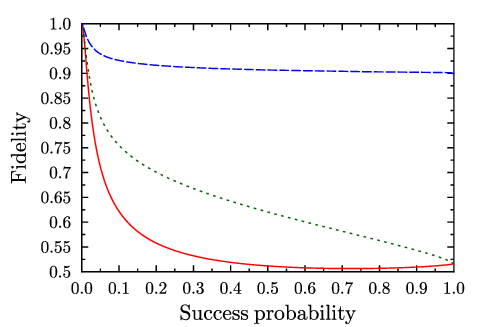}}
	\caption{(color online) Average state fidelity as a function of success probability is plotted for the target state $|\psi_1\rangle=(|0\rangle + |1\rangle)/\sqrt{2}$ and parameters $\kappa=0.5$ and $r=1$.  Red full line: basic acceptance window, green dotted line: advanced acceptance window, blue dashed line: advanced acceptance window and feed-back.}
	\label{fig:fidel11a}
	\end{center}
	\end{figure}

Using the parameters given by Eq. (\ref{eq:multi_roots}), the state obtained by the preparation procedure differs from the desired target state just by  squeezing in the $x_A$ quadrature and can be expressed as $\hat{S}(s)|\psi_{\mathrm{target}}\rangle$, where now $s=\frac{1}{2}\log(\epsilon N +1)$.
As discussed in the previous section, one may employ an anti-squeezing operation that results in rescaling of quadrature operator,
	\begin{equation}
	\label{eq:multi_rescale_x}
	\hat{x}_A \longrightarrow \frac{\hat{x}_A}{\sqrt{\epsilon N+1}}.
	\end{equation}
After such operation the atomic state will become the desired superposition of Dicke states (\ref{eq:objectif}).
Provided that $N\epsilon<1$ one could also pre-squeeze the atomic ensemble before the state engineering procedure and start the protocol from  state $\psi_A(x_A)\propto[-(1-N\epsilon)x_A^2]$. The protocol would then directly yield the required superposition of Dicke states provided that
	\begin{eqnarray}
	\label{eq:multi_roots_sqz}
	  a_{N+1}  &=& R_{N}, \nonumber \\
	 a_{k}  &=& R_{k-1} - R_{k}, \quad k = 2,\ldots, N, \nonumber \\
	a_{1}  &=& - R_{1} - \frac{\epsilon}{1-N\epsilon} \sum_{k=1}^{N} R_k.
	\end{eqnarray}

	\begin{figure}[!t!]
	\begin{center}
	\scalebox{1.0}{\includegraphics{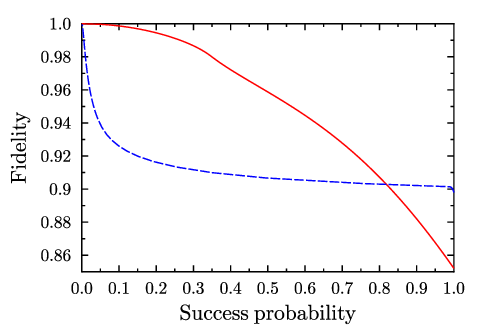}}
	\caption{(color online) Average fidelity is plotted as
	 a function of success probability of preparation for target state 
	 $|\Psi_1\rangle =(|0\rangle + |1\rangle)/\sqrt{2}$, $\kappa=0.5$,  and $r=1$ (blue dashed line, light squeezed in $x_L$) and $r=-1$ (red full line, light squeezed in $p_L$).}
	\label{fig:fidel11b}
	\end{center}
	\end{figure}

\section{Numerical simulations}

The proposal as it has been presented in the previous sections 
assumed conditioning on $p_L=0$. In the experimental realization one can not post-select only cases where $p_L=0$  but has to  use a finite acceptance window to get finite non-zero probability of success. 
The simplest option is to post-select cases, when the result satisfies $|p_L| < \eta$, where $\eta$ is the acceptance threshold. With larger $\eta$ the overall probability of success increases, because the chance of $p_L$ falling into the acceptance window is higher. However as $\eta$ differs from $0$ the prepared state $\rho$ becomes mixed and its fidelity defined as overlap with pure target state decreases.
Thus we obtain a trade-off between fidelity and probability of successful preparation. The  acceptance condition $|p_L|<\eta$ is simple, but not optimal because generally 
fidelity is not a monotonic function of $|p_L|$. One can use a better strategy in selecting which values of $p_L$ are considered successful measurement outcomes. Our advanced acceptance windows are constructed by choosing some fidelity threshold and all values of $p_L$ for which this threshold is exceeded are considered successful outcomes. By changing the fidelity threshold we get the  trade-off between fidelity and success probability, because the lower the fidelity threshold is, the bigger the success probability gets, but the average fidelity gets lower. To improve the resulting fidelity even more, one may employ a feed-back strategy. For each measurement outcome $p_{L}$ it is possible to find corresponding displacement imposed on the atomic state that maximizes the state fidelity.

\begin{figure}[!t!]
	\begin{center}
	\scalebox{0.9}{\includegraphics{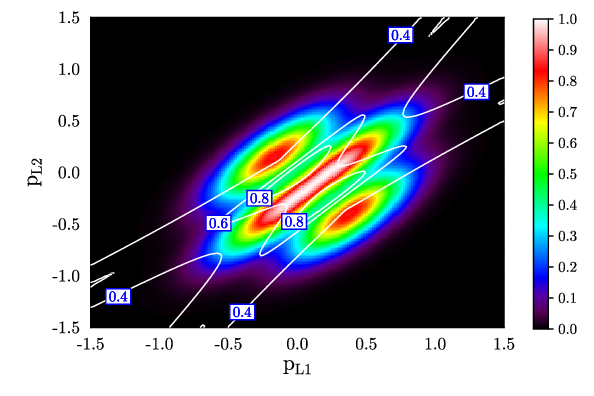}}
	\caption{(color online) Joint probability of measurement outcomes  $p_{L1}$ and 
	$p_{L2}$ in the preparation of Dicke state $|2\rangle$, $\kappa=0.5$, $r=-1$ (colormap).
	The contour lines indicate the fidelity of the prepared state for given measurement outcomes.}
	\label{fig:fidel100mapa}
	\end{center}
	\end{figure}

We present results of numerical  simulation in Fig.~\ref{fig:fidel11a}, where we plot the trade-off between fidelity and success probability in the case of preparation of  $|\psi_1\rangle =(|0\rangle + |1\rangle)/\sqrt{2}$. 
In this simulation we have set $\kappa =0.5$ and $r = 1$. We can see that for this particular target state and parameters, the fidelity of the generated state drops quickly if the basic acceptance condition $|p_L|< \eta$ is used. The advanced acceptance window provides a better
trade-off, as clearly visible in Fig. ~\ref{fig:fidel11a}. Further significant improvement can be achieved if we make use of the feed-back and coherently  displace the state depending on the measurement outcome. Combination of advanced acceptance window and feed-back yields fidelities exceeding $90\%$ even for success probability of the order of $50\%$.

In some situations, it may be advantageous to use  light state squeezed in the phase quadrature $p_L$, hence $r <0$. Fig. \ref{fig:fidel11b} illustrates this effect by comparing the trade-offs between fidelity and success probability obtained for $r=1$ and $r=-1$, respectively, with all other parameters identical. In both cases, we use advanced acceptance windows and optimized feed-back.

The concept of advanced acceptance window can be straightforwardly generalized to preparations of superpositions of first $N+1$ Dicke states, in which case we apply the non-Gaussian operation $\hat{\Theta}_S$ $N$-times and obtain $N$ measurement outcomes $p_{L,j}$. We evaluate the fidelity of the generated state for each set of outcomes $\{p_{L,j}\}$ and the preparation is considered successful if the fidelity exceeds certain threshold. 

\begin{figure}[!t!]
	\begin{center}
	\scalebox{1.0}{\includegraphics{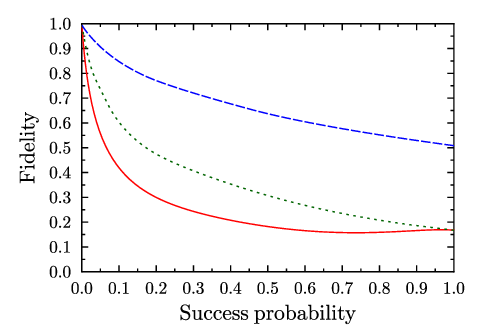}}
	\caption{(color online) Average fidelity as a function of success probability in the process of preparation of $|2\rangle$ state using $\kappa = 0.5, r = -1$. Red full line: basic acceptance window, green dotted line: advanced acceptance window, blue dashed line: advanced acceptance window and feed-back.}
	\label{fig:fidel100}
	\end{center}
	\end{figure}

We demonstrate this procedure on the example of preparation of (squeezed)
Dicke state $|2\rangle$. The protocol involves two operations $\hat{\Theta}_S$ 
interspersed with three displacements and we have two measurement outcomes $p_{L,1}$ and $p_{L,2}$. The probability of detection of a pair of outcomes $p_{L,1}$ and $p_{L,2}$ 
is plotted in Fig.~\ref{fig:fidel100mapa} as a function of $p_{L,1}$ and $p_{L,2}$.
The same figure also contains contour plot of the fidelity of the prepared state 
as a function of $p_{L,1}$  and $p_{L,2}$. Advanced acceptance windows are represented by areas in the plane defined by these fidelity contours.
Note that when calculating the fidelity we again assumed feed-back in form of coherent displacement of the final atomic state and optimized the value of the displacement for each pair of outcomes. The trade-off between fidelity and success probability for the preparation of the  Dicke state $|2\rangle$ is presented in Fig.~\ref{fig:fidel100} which contains comparison of three preparation strategies: basic acceptance window $|p_{L,j}|<\eta$, advanced acceptance window, and advanced acceptance window combined with feed-back.

\section{Complex coefficients}

In previous sections we have described a procedure that is capable of preparation of any superposition of atomic Dicke states with real $R_j$, c.f. Eq. (\ref{eq:multi_fact_form}). In this section we propose a generalized scheme that allows us to prepare any complex superposition. The necessary degrees of freedom are obtained by coupling the light beam to different atomic quadrature at each application of non-Gaussian operation $\hat{\Theta}_S$. Let us consider the rotated atomic quadrature
	\begin{equation}
	\label{eq:rot_quadrature}
	\hat{x}_{A,\phi} = \hat{x}_A \cos\phi +\hat{p}_A\sin\phi.
	\end{equation}
In the experiment, the atoms-light interaction Hamiltonian (\ref{eq:qnd_hamilton}) can be changed to $\hbar \bar{\kappa} \hat{x}_{A,\phi} \hat{x}_L$ by application of magnetic field  which rotates the atomic spin about the $x$ axis. When the light couples to the the rotated atomic quadrature, the interaction operator $\hat\Theta_{S,\phi}$ is given by 
	\begin{equation}
	\label{eq:rot_theta}
	\hat{\Theta}_{S,\phi} \propto \kappa\hat{x}_{A,\phi}\exp\left ( -\epsilon \hat{x}_{A,\phi}^2\right ),
	\end{equation}
	where we assume conditioning on $p_L=0$.
Similarly as before, we combine the $\hat{\Theta}_{S,\phi}$ operator with displacement operators $\hat{D}(\alpha_j)$. Let us now consider only two repeated applications of the $\hat{\Theta}_{S,\phi}$ operator enveloped by three displacement operators. The overall operation reads
	\begin{equation}
	\label{eq:rot_2theta+disp}
	\hat{D}(\alpha_3) \hat{\Theta}_{S,\phi_2} \hat{D}(\alpha_2) \hat{\Theta}_{S,\phi_1} \hat{D}(\alpha_1).
	\end{equation}

	For the sake of simplicity we will assume the limit of strong squeezing of the light beam and/or weak atoms-light coupling, $\epsilon\rightarrow 0$. Under the condition $\alpha_1+\alpha_2+\alpha_3=0$ the action of operator (\ref{eq:rot_2theta+disp}) on atomic vacuum state produces a complex superposition of three lowest atomic Dicke states,
	\begin{equation}
	\label{eq:rot_3dicke}
	|\psi_{A}\rangle \propto |2\rangle + c_1 |1\rangle + c_0 |0\rangle,
	\end{equation}
where $c_1$ and $c_0$ represent relative amplitudes between states $|1\rangle$ and $|2\rangle$, and $|0\rangle$ and $|2\rangle$, respectively.  The two complex numbers $c_0$ and $c_1$ uniquely specify any superposition of states $|0\rangle$, $|1\rangle$ and $|2\rangle$ with non-zero contribution of the highest Dicke state $|2\rangle$.
After some algebra we arrive at the following set of equations for the complex parameters $c_j$,
	\begin{eqnarray}
	\label{eq:rot_c-eq}
	c_0 & = &\frac{1}{\sqrt{2}} (x_1e^{-i\phi_1} x_2 e^{-i\phi_2} + \e^{-2i\phi_2}), \nonumber \\
	c_1 & = & -\frac{1}{\sqrt{2}}(x_1\e^{-i\phi_1} +x_2\e^{-i\phi_2}), 
	\end{eqnarray}
where $x_1 = \Re \left [(\alpha_2 + \alpha_3) \e^{-i\phi_1}\right ]$ and  $x_2 = \Re \left [\alpha_3 \e^{-i\phi_2}\right ]$. This set of equations can be effectively reduced to equation 
for a single complex parameter $z = x_2 \exp{(-i\phi_2)}$,
	\begin{equation}
	\label{eq:rot_z-eq}
	z^2\left (1-\frac{1}{|z|^2}\right) + \sqrt{2}c_1z + \sqrt{2}c_0 = 0.
	\end{equation}
We can formally solve Eq. (\ref{eq:rot_z-eq}) as quadratic equation in $z$ to obtain
	\begin{equation}
	\label{eq:rot_z-sol}
	z = \frac{1}{\sqrt{2}\left (1 - \frac{1}{|z|^2}\right)} \left (-c_1 - \sqrt{c_1^2 - 2\sqrt{2}c_0\left (1 - \frac{1}{|z|^2}\right )}\right ).
	\end{equation}
Taking the absolute values of both sides of Eq. (\ref{eq:rot_z-sol}) we derive an equation for $|z|$,
	\begin{equation}
	\label{eq:rot_modz-sol}
	|z| = \left | \frac{1}{\sqrt{2}\left (1 - \frac{1}{|z|^2}\right)} \left (c_1 + \sqrt{c_1^2 - 2\sqrt{2}c_0\left (1 - \frac{1}{|z|^2}\right )}\right ) \right |.
	\end{equation}
On the left-hand side we have a monotonically  growing function satisfying $\lim_{z\rightarrow \infty}|z|=\infty$. On the right-hand side we have a function that approaches asymptotically a finite real constant for large values of $|z|$ while for $|z|\rightarrow 1$ it grows to infinity. In the interval $(1;\infty)$ the function on right-hand side
of Eq. (\ref{eq:rot_modz-sol}) is continuous. Therefore  there has to exist an intersection point between the graphs of functions on right-hand and left-hand sides. This point is thus a positive real solution of the equation (\ref{eq:rot_modz-sol}).

We have proved that there is always a solution of the equation (\ref{eq:rot_modz-sol}). Moreover, this solution can be determined analytically because  this equation can be 
transformed to a polynomial equation of the fourth degree,
	\begin{eqnarray}
	\label{eq:rot_y-eq}
	d_4y^4 + d_3y^3 + d_2y^2 + d_1y + d_0 = 0,
	\end{eqnarray}
where $y = |z|^2$ and
	\begin{eqnarray}
	\label{eq:rot_y-coef}
	d_4 & = & 1, \nonumber \\
	d_3 & = & -4 - 2|c_1|^2, \nonumber \\
	d_2 & = & 6 + 4|c_1|^2 - 4|c_0|^2 + 2\sqrt{2} (c_1^2c_0^{*} + c_1^{2*}c_0), \nonumber \\
	d_1 & = & -4 - 2|c_1|^2 - 4|c_0|^2|c_1|^2 + 8|c_0|^2, 
	\nonumber \\ & &- 2\sqrt{2} (c_1^2c_0^* + c_1^{2*}c_0) \nonumber \\
	d_0 & = & 1 + 4|c_0|^4 - 4|c_0|^2.
	\end{eqnarray}
Finding the non-negative real root of this polynomial gives the exact solution of the whole problem. We have confirmed numerically for a large number of randomly generated $c_0$ and $c_1$ that there is always such root.
The presented method allows us to prepare arbitrary complex superpositions of the first three atomic Dicke states and overcomes the limitations of the scheme discussed in previous Sections. This technique can be straightforwardly extended to generation of arbitrary superpositions of the first $N$ Dicke states and the effect of finite $\epsilon$ can also be included. However, the resulting nonlinear equations for the coherent displacements $\alpha_k$ and the phase shifts $\phi_j$  become highly complicated and can be solved only numerically.

\section{Direct mapping approach}

In this section we present an alternative strategy for preparation of superpositions of atomic Dicke states. This approach is based on the preparation of specific state of light beam, which is then imprinted into the atomic memory. By using appropriate superposition of the first $N$ Fock states of light and conditioning on $p_L=0$, we can prepare the atoms in a single step in arbitrary desired (squeezed) superposition of the first $N$ Dicke states.

	\begin{figure}[!t!]
	\begin{center}
	\scalebox{1.0}{\includegraphics{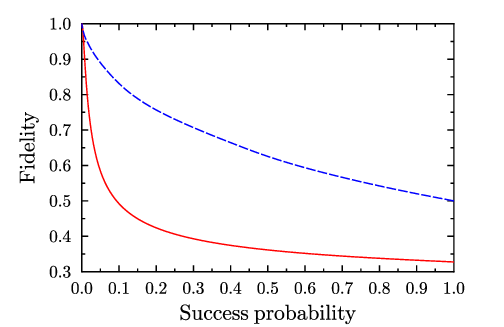}}
	\caption{(color online) Trade-off between fidelity and success probability in the case of preparation of $|2\rangle$ Dicke state. The Fourier approach is indicated by full red line, the original sequential approach by blue dashed line, $\kappa=0.5$.}
	\label{fig:fidel100alt}
	\end{center}
	\end{figure}

The wave-function of target atomic state squeezed by a factor of $\sqrt{1+\kappa^2}$ can be expressed as
	\begin{eqnarray}
	\label{eq:f_target_sq}
	\psi_{A}(x_A) &=&\left[\frac{2(1+\kappa^2)}{\pi}\right]^{1/4} 
	\nonumber \\
	 & & \times \sum_{n=0}^{N} c_n\frac{
	e^{-(1+\kappa^2)x_A^2}
	}{\sqrt{2^n n!}} H_n\left(\sqrt{2(1+\kappa^2)}x_A\right). \nonumber \\
	\end{eqnarray}
Since this wave-function should be obtained by applying the operation $\Phi(\kappa \hat{x}_A)$ onto the vacuum atomic state, we can determine the required wave function of light $\Phi(p)$ in the $p$-representation,
\begin{equation}
\Phi(\kappa x_A)\propto \left(\frac{\pi}{2}\right)^{1/4} e^{x_A^2} \psi_A(x_A).
\end{equation}
Explicitly, we have
\[
\Phi(p) \propto \sum_{n=0}^N c_n \frac{1}{\sqrt{2^n n!}} H_n\left(\frac{\sqrt{2(1+\kappa^2)}}{\kappa}p\right)e^{-p^2}.
\]
This wave function describes a superposition of the first $N+1$ Fock states, 
\begin{equation}
\label{eq:phi_light}
|\phi_L\rangle=\sum_{m=0}^N u_m|m\rangle.
\end{equation}
The coefficients $u_m$ can be determined by calculating the scalar product $\langle m|\phi_L\rangle$ in the $p$-representation. We arrive at the following formula for the non-normalized $u_m$,
\begin{eqnarray}
u_m&=&\sum_{n=0}^N \int_{-\infty}^{\infty}\frac{ i^mc_n }{\sqrt{2^{m+n}\,m! \,n!}} \, H_m(\sqrt{2}p) \nonumber \\
&& \times H_n\left(\frac{\sqrt{2(1+\kappa^2)}}{\kappa}p\right)\, e^{-2p^2} dp. 
\end{eqnarray}
As an example, we have determined the light state $|\phi_L\rangle$ that is required for the preparation of the (squeezed) atomic Dicke state $|2\rangle$.  Explicit calculation yields
	\begin{equation}
	\label{eq:f_light4_2}
	|\phi_L\rangle \propto |0\rangle - \sqrt{2}(1+\kappa^2) |2\rangle.
	\end{equation}
The advantage of the direct single-step preparation is that only one interaction between atoms and light is needed. Moreover, we can generate an arbitrary atomic state by employing an appropriate state of the light beam (\ref{eq:phi_light}). The latter can be prepared from  initial vacuum or squeezed state by combination of coherent displacements and single-photon additions or subtractions \cite{Dakna99,Fiurasek05}. However, the probability of success of such conditional preparation schemes typically decreases exponentially with the maximum number of photons $N$ in the superposition. In contrast, if we use the sequential preparation scheme described in previous sections, then we can wait for successful preparation of the photon-subtracted squeezed state before we switch on the coupling between atoms and light. We still have to condition on the outcomes of the homodyne detections, so even for this  scheme the success rate will decrease with growing $N$ and fixed fidelity of the generated state. Which strategy is optimal thus depends on the particular target state and other experimental parameters.

Figure~\ref{fig:fidel100alt} shows the trade-off between fidelity and success probability in the case of preparation of the $|2\rangle$ Dicke state. This trade-off is compared with the trade-off for the original sequential strategy. Note that the success probability represents only the probability of acceptance of the measurement outcomes $p_L$ of homodyne detector and does not include the probability of generation of the required states of light. As discussed above, the total cost of preparation of the  non-Gaussian states of light would be lower in case of the sequential strategy than in case of the single-step strategy.  In the numerical simulations, we assumed advanced acceptance window and optimized  feed-back on the atomic state.
 Interestingly, the sequential strategy outperforms the direct strategy in this case as can be seen in Fig.~\ref{fig:fidel100alt}.


\section{Conclusions}

In this paper, we have proposed a protocol capable of preparation of a wide class of highly non-classical states of atomic quantum memory. We have shown that a non-Gaussian operation on the state stored in memory can be performed using non-Gaussian state of light, QND coupling between atoms and light and conditioning on the homodyne detection performed on the output light beam. Based on this elementary non-Gaussian operation, we have devised a simple strategy for preparation of real superpositions of atomic Dicke states by sequential application of coherent displacement and non-Gaussian operation $\hat{\Theta}_S$. Several numerical simulations have been performed to verify the functionality of our scheme. We have found that the performance of the protocol can be significantly improved by judicious choice of the acceptance window for results of homodyne detection $p_L$ and by applying feedback displacement on atoms depending on $p_L$. We have also shown that using coupling of light to arbitrary rotated atomic quadratures allows us to prepare any complex superpositions. Finally, we have presented a general single-step scheme for preparation of atomic Dicke state based on imprinting a state of light beam onto quantum memory. 
We hope that the obtained results will stimulate attempts to experimentally generate highly non-classical states of atomic memory and that they will be found useful in development of advanced quantum information processing schemes with atomic memories requiring non-Gaussian operations.

\acknowledgments
We acknowledge financial support from the Ministry of Education of the Czech Republic
under the research projects Center of Modern Optics (LC06007), Research centre - Optical structures, detection systems and related technologies for low photon number applications (1M06002) and Measurement and Information in Optics (MSM6198959213).
We also acknowledge the financial support of the Future and Emerging Technologies (FET) programme within the Seventh Framework Programme for Research of the European Commission, under the FET-Open grant agreement COMPAS, number 212008 and co-funding of this project by MSMT (7E08028).


\begin{thebibliography}{10}

\bibitem{Scarani08}
V. Scarani, H. Bechmann-Pasquinucci, N.J. Cerf, M. Dusek, N. Lutkenhaus, and M. Peev,
arXiv:0802.4155. 

\bibitem{Briegel98}
H.-J. Briegel, W.~D{\"u}r, J.~I. Cirac, P.~Zoller,
Phys. Rev. Lett. \textbf{81}, 5932 (1998).

\bibitem{Duan01}
L.~M. Duan, M.~D. Lukin, J.~I. Cirac, P.~Zoller,
Nature (London) \textbf{414}, 413 (2001).



\bibitem{Matsukevich04}
D.~N. Matsukevich, A.~Kuzmich,
Science \textbf{306},  663 (2004).


\bibitem{Chaneliere05}
T. Chaneliere, D. Matsukevich, S. D. Jenkins, S.-Y. Lan, T.A.B. Kennedy, and A. Kuzmich, 
Nature \textbf{438}, 833 (2005).

    

\bibitem{Choi08}
K. S. Choi, H. Deng, J. Laurat, and H. J. Kimble, Nature \textbf{452}, 67 (2008).


\bibitem{Yuan08}
Z.-S. Yuan, Y.-A. Chen, B. Zhao, S. Chen, J. Schmiedmayer, J.-W. Pan,
Nature \textbf{454}, 1098 (2008). 



\bibitem{Smith08}
 M. H. Schleier-Smith, I. D. Leroux, V. Vuletic, arXiv:0810.2582.
 


\bibitem{Appel08}
J. Appel, P.J. Windpassinger, D. Oblak, U.B. Hoff, N. Kjaergaard, E.S. Polzik, arXiv:0810.3545.
    
    
\bibitem{Julsgaard01} 
B. Julsgaard, A. Kozhekin, and E.S. Polzik,
Nature (London) \textbf{413}, 400 (2001). 


\bibitem{Schori02}
C.~Schori, B.~Julsgaard, J.~L. S{\o}{}rensen, E.~S. Polzik,
Phys. Rev. Lett. \textbf{89}, 057903 (2002).


\bibitem{Julsgaard04}
B. Julsgaard, J. Sherson, J. I. Cirac, J. Fiurasek, and E. S. Polzik, 
Nature (London) \textbf{432}, 482 (2004).


 \bibitem{Cviklinski08}
J. Cviklinski, J. Ortalo, J. Laurat, A. Bramati, M. Pinard, and E. Giacobino, 
Phys. Rev. Lett. \textbf{101}, 133601 (2008).
      

\bibitem{Kuzmich98}
A. Kuzmich, N.P. Bigelow, and L. Mandel,
Europhys. Lett. \textbf{42}, 481 (1998). 


\bibitem{Duan00}
L.-M. Duan, J. I. Cirac, P. Zoller, and E. S. Polzik, 
Phys. Rev. Lett. \textbf{85}, 5643 (2000).


\bibitem{Kuzmich00}
A.~Kuzmich, E.~S. Polzik,
Phys. Rev. Lett. \textbf{85}  (2000) 5639


\bibitem{Hammerer08}
K. Hammerer, A.S. Sorensen, E.S. Polzik,  arXiv:0807.3358.
    
    
\bibitem{Dantan05b}
A.~Dantan, A.~Bramati, M.~Pinard,
Phys. Rev. A \textbf{71}, 043801 (2005).

    
\bibitem{Sherson06}
J.F. Sherson, H. Krauter, R.K. Olsson, B. Julsgaard, K. Hammerer, I. Cirac, and E.S. Polzik, Nature (London) \textbf{443}, 557 (2006). 



\bibitem{Fiurasek03}
J. Fiur\'{a}\v{s}ek, Phys. Rev. A \textbf{68}, 022304 (2003). 


\bibitem{Molmer04} 
K. Molmer and L. B. Madsen, Phys. Rev. A \textbf{70}, 052102 (2004). 


\bibitem{Hammerer05}
K. Hammerer, E.S. Polzik, and J.I. Cirac,
Phys. Rev. A \textbf{72}, 052313 (2005). 


\bibitem{Muschik06}
C.A. Muschik, K. Hammerer, E. S. Polzik, and J.I. Cirac,
Phys. Rev. A \textbf{73}, 062329 (2006). 


\bibitem{Sherson06b}
J. Sherson, A.S. Sorensen, J. Fiur\'{a}\v{s}ek, K. Molmer, and E.S. Polzik, 
Phys. Rev. A \textbf{74}, 011802(R) (2006). 


\bibitem{Fiurasek06}
J.~Fiurasek, J.~Sherson, T.~Opatrny, E.~S. Polzik,
Phys. Rev. A \textbf{73}, 022331  (2006).


\bibitem{Eisert02}
J. Eisert, S. Scheel, and M.B. Plenio, 
Phys. Rev. Lett. \textbf{89}, 137903 (2002).

\bibitem{Giedke02} 
G. Giedke, and J.I. Cirac, Phys. Rev. A \textbf{66}, 032316 (2002).

\bibitem{Fiurasek02} 
J. Fiurasek, Phys. Rev. Lett. \textbf{89}, 137904 (2002).


\bibitem{Browne03}
D. E. Browne, J. Eisert, S. Scheel, and M. B. Plenio,
 Phys. Rev. A \textbf{67}, 062320 (2003).


\bibitem{Eisert04}
J. Eisert, D.~E. Browne, S. Scheel, and M.~B. Plenio,
Ann. Phys. \textbf{311}, 431 (2004).


\bibitem{Hage08}
B. Hage, A. Samblowski, J. DiGuglielmo, A. Franzen, J. Fiur\'{s}\v{s}ek, and R. Schnabel,  
Nature Phys. \textbf{4}, 915 (2008).


\bibitem{Dong08}
R. Dong, M. Lassen, J. Heersink, C. Marquardt, R. Filip, G. Leuchs, and U.L. Andersen,
Nature Phys. \textbf{4}, 919 (2008).



\bibitem{Massar03}
S. Massar and E. S. Polzik, Phys. Rev. Lett. \textbf{91}, 060401 (2003).


\bibitem{Filip08}
R. Filip, Phys. Rev. A \textbf{78}, 012329 (2008).


\bibitem{Ourjoumtsev06}
A. Ourjoumtsev, R. Tualle-Brouri, J. Laurat, and Ph. Grangier,
Science \textbf{312}, 83 (2006). 


\bibitem{Nielsen06} 
J.S. Neergaard-Nielsen, B.M. Nielsen, C. Hettich, K. Molmer, and E.S. Polzik, 
Phys. Rev. Lett. \textbf{97}, 083604 (2006).


\bibitem{Wakui06} 
K. Wakui, H. Takahashi, A. Furusawa, and M. Sasaki, 
Opt. Express \textbf{15}, 3568 (2007).



\bibitem{Dakna99}
M. Dakna, J. Clausen, L. Knoll, and D.-G. Welsch, Phys.
Rev. A \textbf{59}, 1658 (1999); Phys. Rev. A \textbf{60}, 726 (1999).

\bibitem{Fiurasek05}
J. Fiurasek, R. Garcia-Patron, and N.J. Cerf,
Phys. Rev. A \textbf{72}, 033822 (2005). 



\end{thebibliography}
\end{document}